\documentclass[useAMS,usenatbib,usegraphicx]{mn2e}

\newcommand\aj{AJ}
\newcommand\apj{ApJ}
\newcommand\apjs{ApJS}
\newcommand\aap{A\&A}
\newcommand\aaps{A\&AS}
\newcommand\mnras{MNRAS}

\title[]{The cluster galaxy luminosity function at $z=0.3$: a
recent origin for the faint-end upturn ?}
\author[D. Harsono and R. De Propris]
{
D. Harsono$^{1}$ and
R. De Propris$^{2}$\thanks{E-mail: rdepropris@ctio.noao.edu}\\
$^{1}$ Department of Physics and Astronomy, University of California at
Los Angeles, USA\\
$^{2}$ Cerro Tololo Inter-American Observatory, Casilla 603, La Serena, Chile
}

\begin{document}

\date{}

\pagerange{\pageref{firstpage}--\pageref{lastpage}} \pubyear{2007}

\maketitle

\label{firstpage}

\begin{abstract}

We derive deep luminosity functions (to $M_z=-15$) for galaxies in
Abell 1835 ($z=0.25$) and AC 114 ($z=0.31$) and compare these with
the local $z'$ luminosity function for 69 clusters. The data show
that the faint-end upturn, the excess of galaxies above a single
Schechter function at $M_z < -17$, does {\it not} exist in the higher
redshift clusters. This suggests that the faint-end upturn galaxies
have been created recently, by infall into clusters of star-forming
field populations or via tidal disruption of brighter objects.

\end{abstract}


\section{Introduction}

The mass function of Cold Dark Matter (CDM) halos emerging from the epoch
of recombination is expected to be very steep, with a slope $\alpha
\sim -2$ \citep{klypin99,moore99}. However, the slope of the luminosity 
function (LF) of galaxies in the Local Group is very flat, $\alpha \sim -1.1$
to $M_V = -10$ \citep{pritchet99}, while Trentham, Sampson \& Banerij (2005)
also find a very flat slope (to $M_g=-9$) in poor groups selected from
the Sloan survey. This is the well known small scale problem in CDM, for
which a number of solutions (involving suppression of dwarf galaxy formation
in small haloes) have been proposed. On the other hand, there is evidence 
that the LF of dwarf galaxies ($M_V < -17$) in rich clusters shows a steep upturn,
approaching the predicted CDM value. Originally discovered by \cite{driver94}
and, independently, by \cite{depropris95}, the existence of this faint-end
upturn has recently been placed on a firmer footing by Popesso et al. (2006, 
hereafter P06), who detect it in all four bands of the composite LF of 69 
X-ray selected  clusters with Sloan photometry. It is very unlikely that 
fluctuations in the background counts would be able to affect all 69 objects 
equally and produce an artificial steepening of the LF at faint luminosities.\\

The origin of the upturn should have interesting consequences for theories of
structure formation. If the upturn galaxies are primordial, they may be the
relic building blocks of the original population of dwarf-sized fragments
that went into constructing the cluster giants. However, the fact that the
upturn is observed in rich environments but not in poor groups suggests that
its origin is related to the cluster environment. Unless the cluster acts
to preserve a primordial population of faint dwarfs \citep{babul92}, it
appears more likely that the upturn galaxies have undergone recent infall
from the surrounding field \citep{wilson97}. They may also have been
whittled down from more massive objects (Conselice, Wyse \& Gallagher 2001;
\citealt{conselice03}). Although the colors of present-day upturn galaxies
are consistent with those of dwarf spheroidals on the red sequence (P06,
\citealt{yamanoi07}), \cite{conselice01} argues that one half of the fainter
Virgo dwarfs are actually blue.

The approach we follow here is to study the evolution of the faint-end
upturn. In practice, we derive deep LFs for distant clusters and attempt
to compare the differential luminosity evolution of the bright cluster
members (which evolve passively) and the upturn galaxies. If the upturn 
galaxies are primordial, they should evolve with the same speed as the
giants, most of whose stellar populations were formed at high redshift.
Conversely, if the upturn consists of fading irregular galaxies \citep{
wilson97}, we should see a brighter onset of the steepening at higher
redshifts, or, if the upturn galaxies have undergone recent infall from
the field, a much flatter faint-end slope.

We begin by considering two clusters at $z=0.3$ with deep archival Hubble
Space Telescope (HST) Advanced Camera for Surveys (ACS) imaging. While
this choice is set by the availability of archival data, this is an 
interesting redshift as it corresponds to the epoch at which we witness
the onset of the faint blue galaxies excess and the increase in the
blue fraction in clusters of galaxies, both of which may be attributed
to a population of star forming low luminosity galaxies temporarily
brightened by star formation episodes (e.g., \citealt{babul96,driver96})

We derive a deep $z$ band LF for galaxies in the above clusters and compare
with the P06 local sample. This latter represents the averaged LF of $z < 0.1$
clusters to a depth comparable to our own and determined using similar
methods (statistical subtraction of non cluster members using counts in
reference fields). An additional benefit is that the ACS filters are
designed to imitate the Sloan passbands used in the photometry of local
clusters by P06. For these reasons, we adopt their $z'$ band LF as our
comparison to measure evolution.

The following section describes the data, their reduction, analysis and photometry.
Discussion of the results and their interpretation can be found in section 3.
We adopt the latest cosmology with $\Omega_M=0.3$, $\Omega_{\Lambda}=0.7$ and
$H_0=70$ km s$^{-1}$ Mpc$^{-1}$. 

\section{Observations and Data Analysis}

We use two deep (10,000s total integration) archival observations of
the clusters Abell 1835 (at $z=0.253$) and AC 114 ($z=0.312$) carried
out with the ACS in the $z$ (F850LP) filter (PI: Pell{\'o}; PID:
10154). Abell 1835 is a richness 0 cluster, similar to the Fornax
or Virgo clusters (but somewhat more massive), while AC 114 has
richness class 2 and is therefore similar to the Coma cluster. Both
clusters are therefore comparable to the P06, although they lie on
the high mass end of the distribution.

We retrieved the individual flatfielded exposures from the HST archive
and processed them through {\tt Multidrizzle} \citep{koekemoer02} to
produce a single registered image for each cluster, interpolated across
the chip gaps and with cosmic rays and cosmetic defects removed. The
final data products are shown in Fig.~1. The images cover a total of
900 kpc on the side. For AC 114 the image is centred on the brightest
cluster galaxies and spans a radius of 400 kpc, which is equivalent to
1/2 of the median $r_{200}$ (the radius at which the cluster is 200
times denser than the field) over which P06 compute their LFs. The geometry
of A1835 is more complex. Because of a bright star, the central cluster
galaxy is shifted towards the eastern side of the image. The ACS image
of this cluster reaches out to the median $r_{200}$ value.

\begin{figure}
\includegraphics[width=8cm]{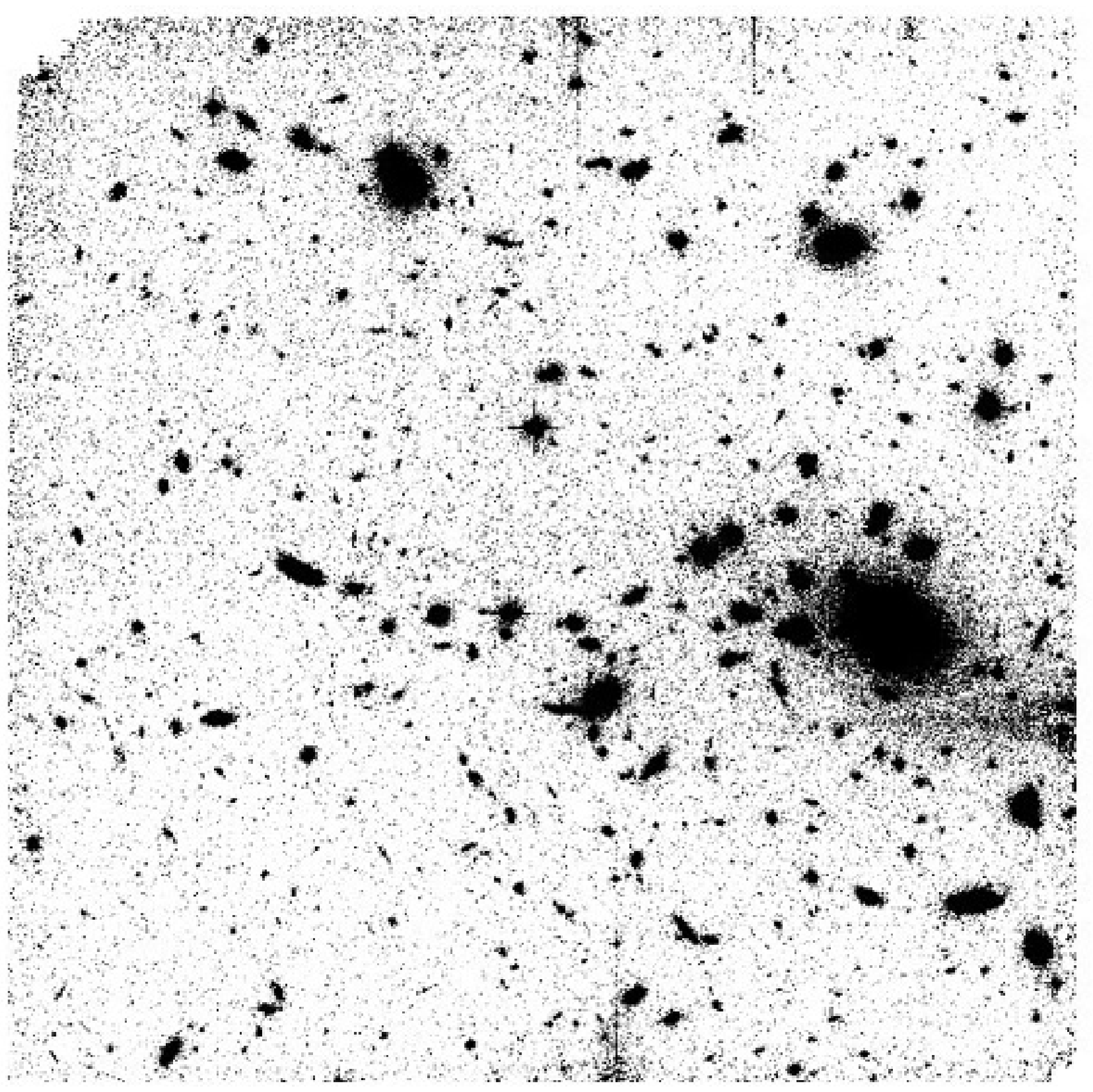}\\
\includegraphics[width=8cm]{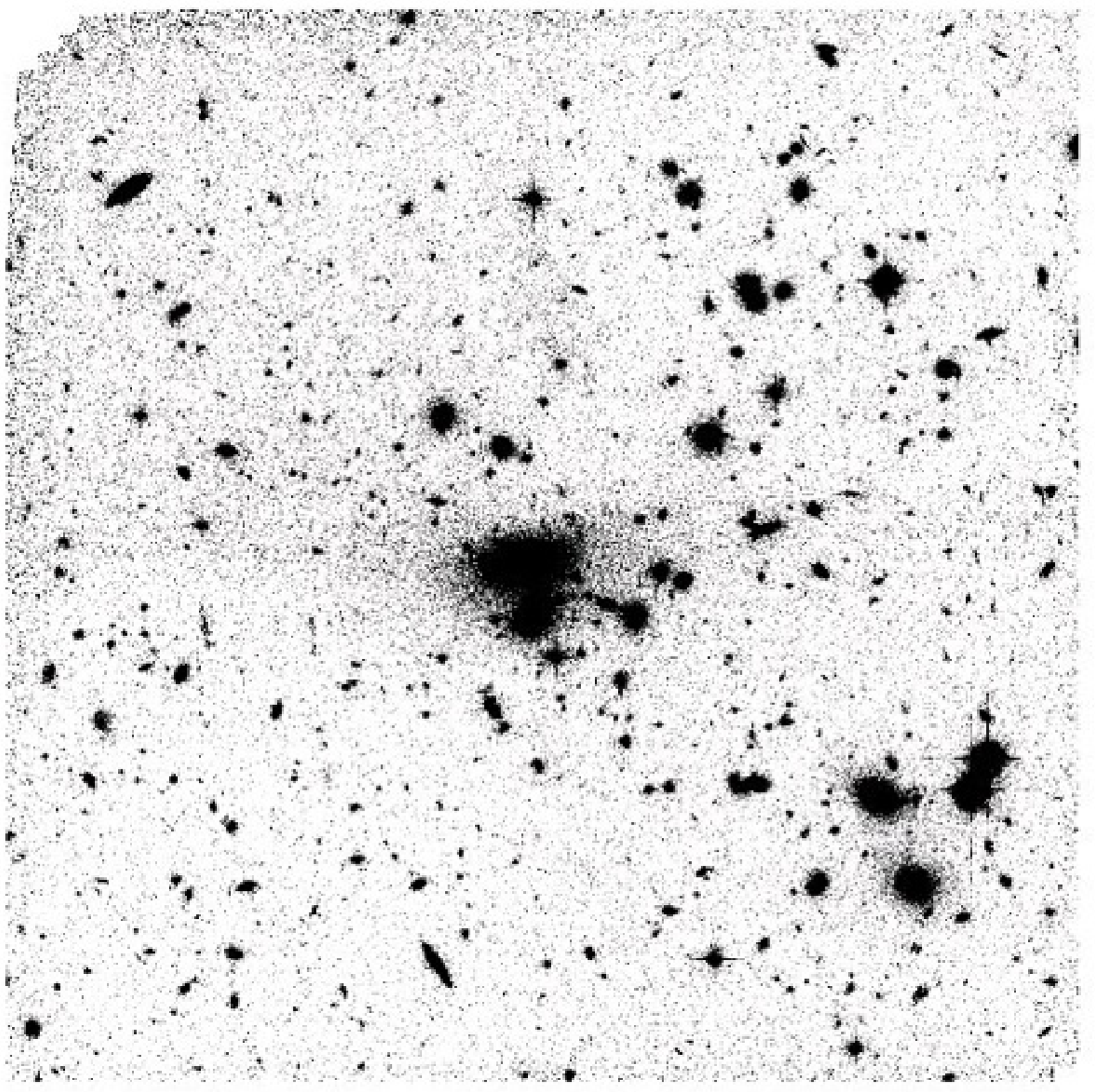}
\caption{Grayscale images of A1835 (top) and AC114 (bottom)}
\end{figure}

As with all similar studies, and especially at the faint end, we can
only determine cluster membership statistically. Counts of field
galaxies projected onto the cluster field of view can be estimated
by observation of blank (cluster-less) fields. We note here that 
the luminosity distance to A1835 and AC114 is much larger than the
largest structure present in local redshift surveys and therefore 
the counts in the direction of the clusters should reflect the
cosmic mean (the structure is uncorrelated with the cluster). We use 
the two GOODS fields \citep{giavalisco04} as our reference fields, 
as these are the deepest and widest available images in the $z$ band.

Detection and photometry were carried out using SExtractor \citep{bertin96}
using exactly the same parameters for the clusters and GOODS fields:
a minimum detection `aperture' of 7 connected pixels above $1.5\sigma$
from the sky, which is equivalent to a 3$\sigma$ detection. We measured
photometry in an adaptive aperture and in a circular aperture of area
equivalent to the minimum number of connected pixels, to derive a mean
central surface brightness. The data were calibrated on the HST AB system 
using published zeropoints.

Fig.~2 shows plots of $\mu_z$ vs $z$ for objects in the cluster fields
and the GOODS fields. The sequence of objects which constitues an upper
envelope to the detections can be identified with stars. We can easily 
separate stars and galaxies to at least $z=25.5$, which is fainter than
the completeness limit we estimate below. We inspected our detections to remove 
a few objects (in both cluster and GOODS fields) that were deblended 
excessively by the Sextractor algorithm (usually bright spiral galaxies). 
For the few objects where this was necessary, we computed magnitudes in a 
single large aperture. 

\begin{figure*}
\begin{tabular}{cc}
\includegraphics[width=8cm]{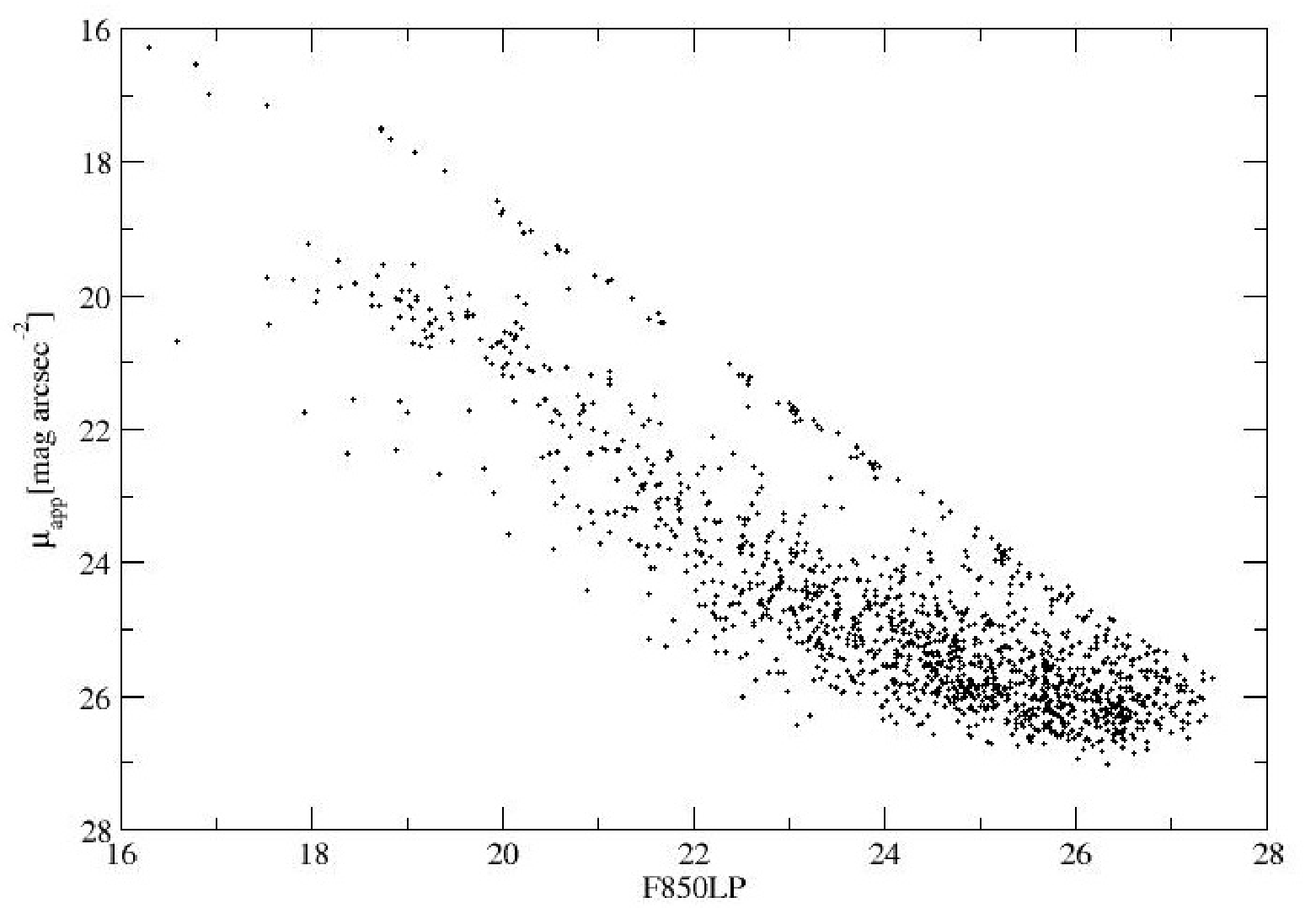} & \includegraphics[width=8cm]{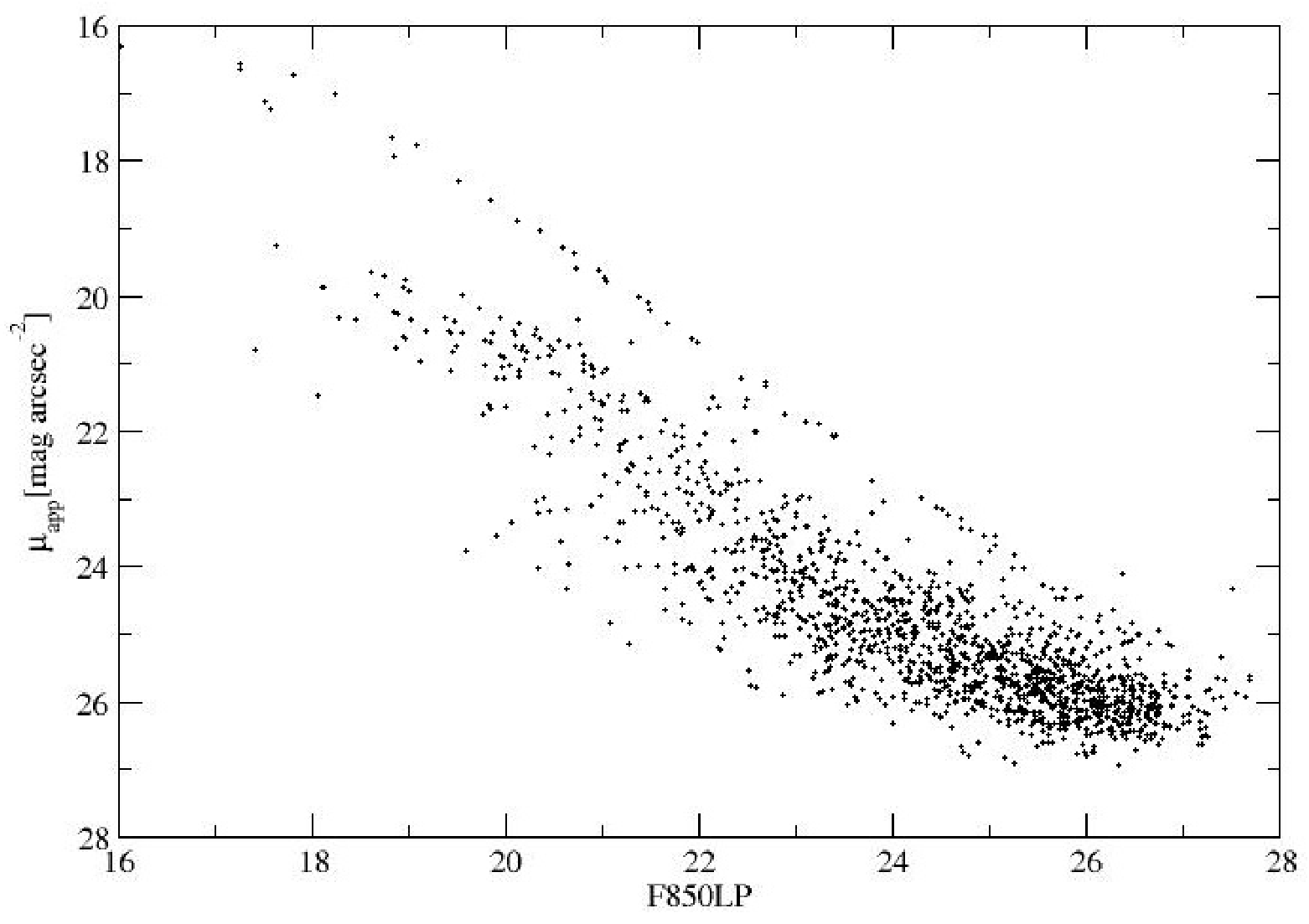}\\
\includegraphics[width=8cm]{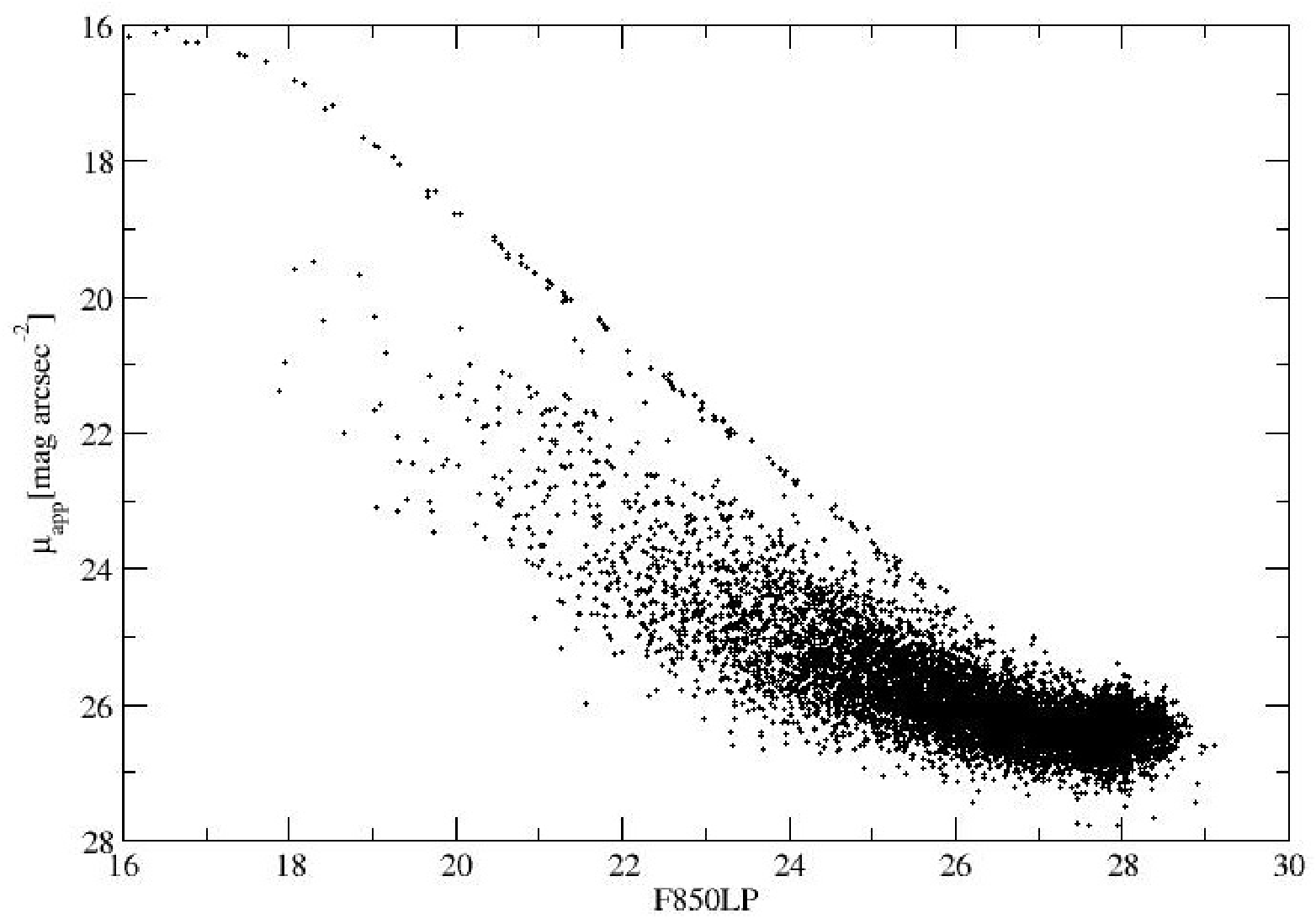} & \includegraphics[width=8cm]{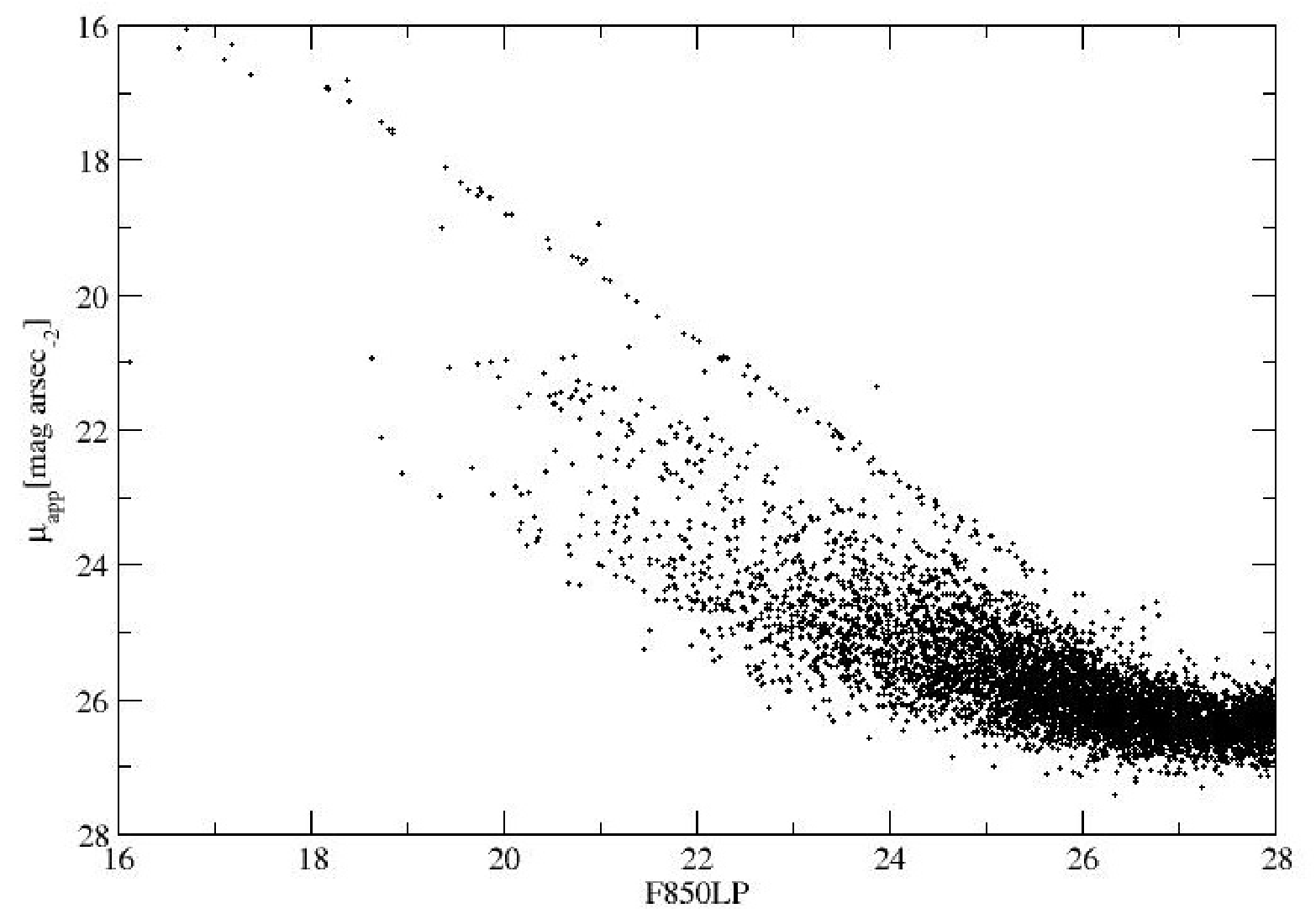}\\
\end{tabular}
\caption{Central surface brightness vs. $z$ for objects in the A1835
(top left), AC114 (top right), GOODS-N (bottom left) and GOODS-S (bottom
right) fields}
\end{figure*}

It is obvious that the GOODS fields reach deeper apparent luminosities
and somewhat deeper surface brightness limits than our cluster field (as
they have about 10 times the exposure time). We need to choose a common 
limiting central surface brightness for both the cluster and the background 
fields. We only count galaxies with central $\mu_z$ above this value. It is 
seen from Fig.~2 that $\mu_z \sim 26.2$ mag arcsec$^{-2}$ selects most objects 
in the cluster fields. We note that we start missing lower surface brightness
objects at $z > 24$ so that we are incomplete at fainter apparent
luminosities. However, we are equally incomplete in the GOODS fields
so this should not affect our determination of the LF, although,
strictly, we can only state a lower limit to the LF slope.

Fig.~3 shows galaxy number counts (with central $\mu_z < 26.2$) for
the cluster and GOODS fields. From this figure, we see that our data
are complete at least to $z=24.5$ and possibly beyond (the counts are
still rising). We adopt this as our magnitude completeness limit.

\begin{figure}
\includegraphics[width=8cm]{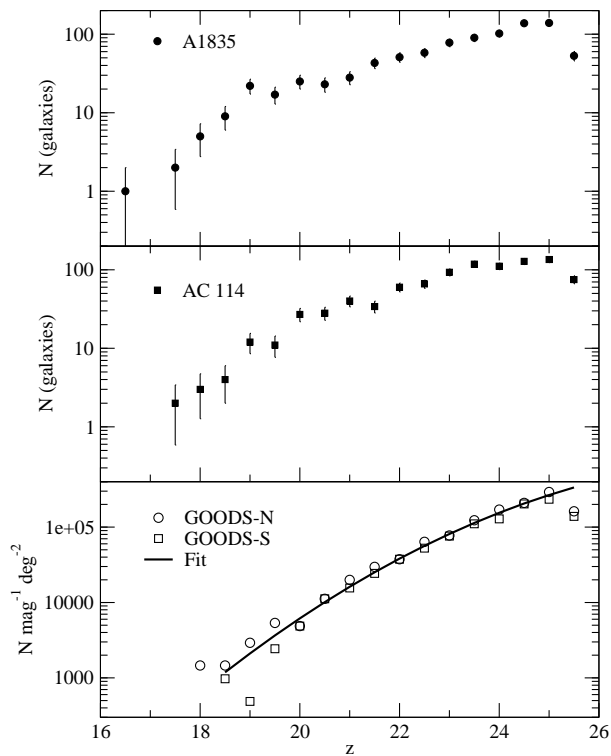}
\caption{Galaxy number counts vs. $z$ for A1835, AC114, GOODS-N and GOODS-S
and the best fitting quadratic to the logarithmic number counts 
$\log_{10} N= -14.256 + 1.3623*z-0.023002*z^2$. The error bars for
counts in the GOODS fields are comparable to the size of the points
and are omitted for clarity.
}
\end{figure}

In order to remove contamination of our cluster fields by foreground and
background galaxies we use galaxy counts in the GOODS fields. We fit these
counts with a second degree polynomial to smooth variations due to large
scale structure. At faint luminosities, the typical scale sampled by each
GOODS field is also larger than the largest existing structure, and the
counts should therefore reflect the cosmic mean. This can be seen most clearly
in Figure 12 of \cite{capak07} where the $I$ band counts for COSMOS fields,
Hubble Deep Field (from \citealt{capak04}) and the HST counts in the Hubble
Deep Fields North and South are well within each other's error bars. Fig.~3
also shows how the GOODS North and South counts are consistent with each
other. We note also that the effect of cosmic variance on the background
counts can be estimated in the error budget, using the `counts-in-cells'
approach of \cite{peebles75}.

We subtract the scaled (and smoothed) GOODS counts from the cluster
counts and compute error statistics by adding, in quadrature, the
Poisson errors in the galaxy counts for the cluster fields, the
scaled errors for the predicted background contribution and the
Poisson errors for counts in the GOODS fields. To these we add,
also in quadrature, the clustering errors (which take care of 
cosmic variance) for galaxy counts in the GOODS fields and the
scaled galaxy counts for the cluster fields, following the methods
of \cite{huang97,driver03} and \cite{pracy04}. For reference, the
error in the counts for field $i$ at apparent magnitude $m$ is computed 
as \citep{huang97}:

$$\sigma_i^2=N_i(m) + 5.3 (r/r_*)^{\gamma} \Omega_i^{(1-\gamma)/2} N^2_i(m)$$

where $N_i(m)$ are the galaxy counts for fields $i$ at apparent magnitude $m$,
$r_*$ is defined as $5 \log r_*=m -M^*-25$ ($M^*$ is as defined in the Schechter
function), $\Omega_i$ is the area of field $i$ and $\gamma$ is the index of
the correlation function ($1.77$).

\section{Discussion}

We plot the LFs and best fits (to a single Schechter function) to
the galaxy counts in A1835 and AC114 in Fig.~4. We see that a single
Schechter function provides a reasonable fit to the data. The best
fit parameters (with marginalised $1\sigma$ errors) are: $M_z^*=17.85
\pm 0.97$ and $\alpha=-1.38 \pm 0.13$ for A1835 and $M_z^*=18.99
\pm 0.86$ and $\alpha=-1.30 \pm 0.12$ for AC114. The characteristic
luminosity is poorly determined, because of small number statistics,
but is consistent with the local value of P06 after $k+e$
correction (using a \citealt{bruzual03} model with solar metallicity, Salpeter 
initial mass function, formation redshift of 3 and e-folding time of 1 Gyr, 
which provides a good fit to the colours of present-day ellipticals) and
correcting for the $+0.52$ mag. offset between HST $z$ and SDSS $z'$ \citep{holberg06}. 
The slope we derive is somewhat steeper than the local one but is consistent 
with the slope of the LF in A2218 measured by \cite{pracy04}: rich clusters are
known to contain more dwarfs and have steeper faint-end slopes \citep{phillipps98}.

\begin{figure*}
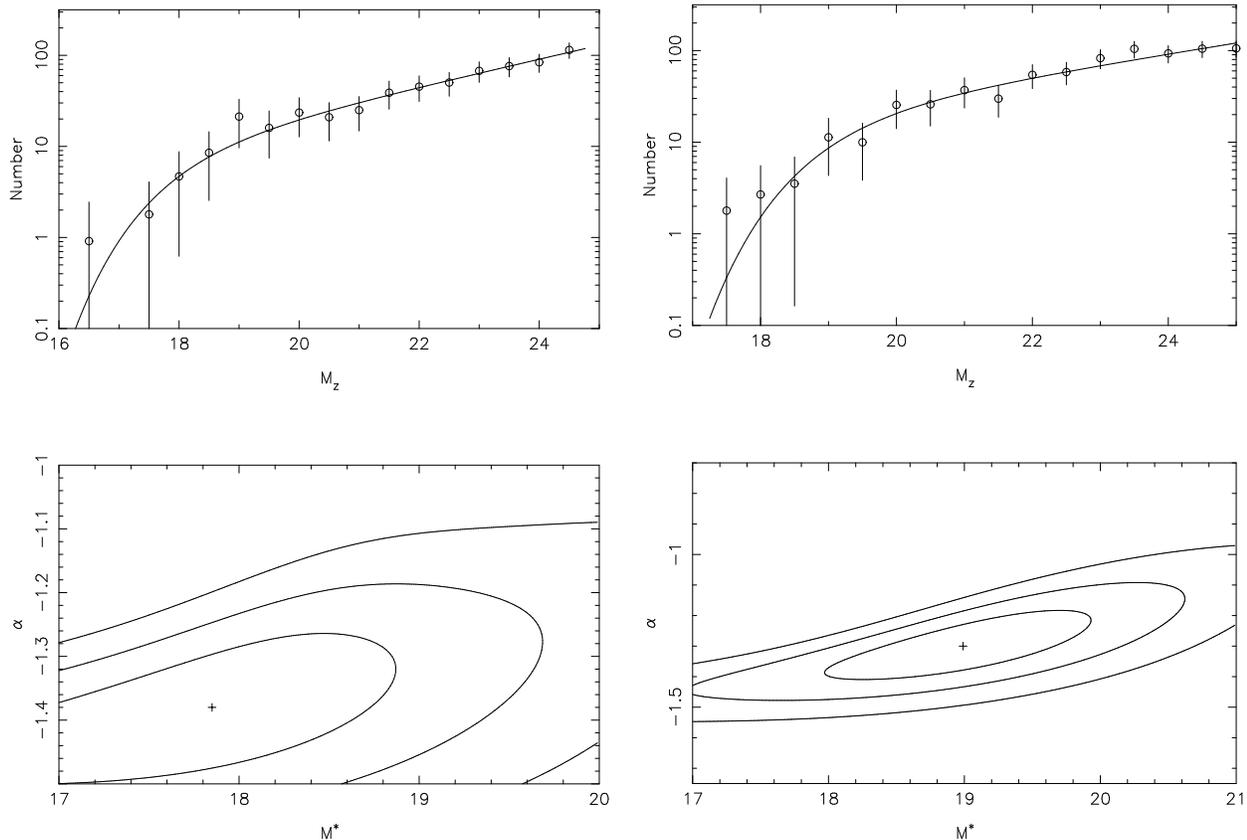

\begin{tabular}{cc}
\includegraphics[width=8cm]{fig6.ps} & \includegraphics[width=8cm]{fig7.ps}\\
\end{tabular}
\caption{Luminosity functions and best fits, with error ellipses, for
A1835 (left) and AC114 (right)}
\vspace{1cm}
\end{figure*}

Our main interest is to compare the local LFs with our observations
in order to study the evolution of the faint-end upturn. To this end
we plot the LFs by P06 over our data, after correcting
for the distance modulus, $k+e$ correction and magnitude offset.
The normalization is chosen to match our counts at $M_z=-21$ in
order to better compare the faint-end behaviour. It is obvious that
there is {\it no} evidence of a faint-end upturn at $M_z < -17$
in the $z=0.3$ clusters.

\begin{figure*}
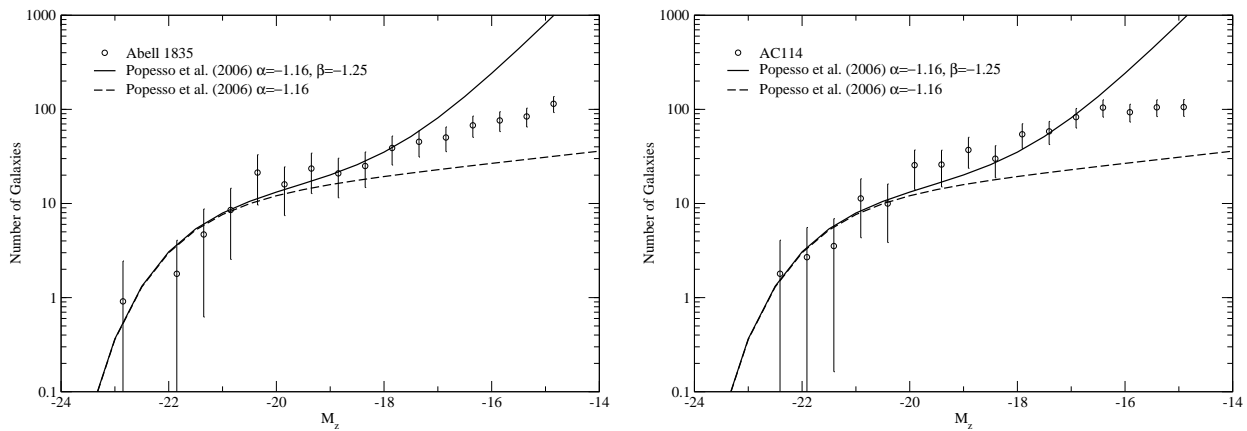

\begin{tabular}{cc}
\includegraphics[width=8cm]{fig8.ps} & \includegraphics[width=8cm]{fig9.ps}\\
\end{tabular}
\caption{Comparison of local LFs to the A1835 (top) and AC114 (bottom)
LFs. We also show an upturn-less LF for comparison.}
\end{figure*}

It may be argued that the upturn is most evident in the outskirts of
clusters, rather than in their cores (P06). \cite{pracy04} find that
the LF of what they term `ultra-dwarfs' is flatter in the core than
in the outskirts. However, we sample between 1/2 of $r_{200}$ and 
$r_{200}$ and P06 detect a faint-end upturn, at least for red galaxies 
to which we are most sensitive in the $z$ band, even for $r < 0.3 r_{200}$ 
(their Figure 11). Furthermore, the upturn is clearly detected by 
numerous authors even in the core (indeed, especially in) of the 
Coma cluster \citep{depropris98,trentham98,milne07}.

The two clusters we survey are also similar to local clusters: they have
the same richness as Coma or Fornax and have X-ray properties similar to
the sample of P06. It is unlikely that differences in the cluster samples 
are responsible for our findings.

Neither does it appear likely that we are missing galaxies (with respect
to the SDSS) because of low surface brightness effects: SDSS data
are 50\% complete at $\mu_r=23.5$ \citep{blanton05} while we go considerably
deeper (at least $\mu_z=26.2$). In AC114 and AC118 Andreon, Punzi \&
Grado (2005) present a $K$-band LF of depth comparable to ours. The
parameters for the LF in AC114 are virtually identical to ours and
both LFs can be fitted by a single Schechter function with no upturn.

The most obvious interpretation is that the faint-end upturn is of
recent origin. One possibility is that it consists of fading objects
that have recently infallen from the field, where the LF for 
star-forming galaxies is actually quite steep \citep{hogg97}. These
galaxies may also be whittled down from formerly more massive
objects, in the manner proposed by \cite{conselice01}. It is interesting to
speculate that these faint dwarfs are but the latest instalment
in the buildup of the cluster red sequence. As we look back in time
we see that the cluster dwarf population is progressively missing
as low mass field galaxies are converted into dwarf ellipticals
(e.g., \citealt{delucia07}): these very faint dwarfs may correspond to
the very low amplitude fluctuations which are nowadays falling into
the large scale structures for the first time.

Ultimately, we should be able to achieve a more complete understanding
of these objects by studying a large number of clusters in several
bandpasses and over wide fields. Some HST archival data for this
project are available and are being analyzed for a future publication
but these include only the central regions of several intermediate
redshift clusters. Wide-field imaging on large telescopes may be 
needed to make further progress

\section*{Acknowledgements}

DH carried out this work while on the Research Experience
for Undergraduate program at Cerro Tololo Inter-American
Observatory and wishes to acknowledge the hospitality of
CTIO  during the Chilean summer. The REU program is supported
by NSF grant 0353843. We would like to thank the CTIO REU
program director, Stella Kafka, for her efforts on behalf of
the students.

\label{lastpage}

\end{document}